\documentclass[12pt]{article}

\usepackage{epsfig}
\usepackage{amsfonts}

\usepackage{dsfont}

\usepackage{times}

\def\+{{+\!\!\!+}}

\def\bw{{\bar w}}
\newcommand{\C}{{\mathds C}}

\def\L{{\cal L}}
\def\K{{\cal K}}

\def\F{{\cal F}}

\def\bR{\hbox{R\hspace{-0.10in}I\hspace{0.04in}}} 

\def\pmb#1{\setbox0=\hbox{#1}%
\kern.0em\copy0\kern-\wd0
\kern-.04em\copy0\kern-\wd0
\kern.08em\copy0\kern-\wd0
\kern-.04em\raise.0433em\box0 }         

\newcommand{\nc}{\newcommand}
\nc{\beq}{\begin{equation}}
\nc{\eeq}[1]{\label{#1}\end{equation}}
\nc{\ber}{\begin{eqnarray}}
\nc{\eer}[1]{\label{#1}\end{eqnarray}}
\nc{\pek}[1]{\cite{#1}}
\nc{\enr}[1]{(\ref{#1})}
\nc{\kal}[1]{{\cal{#1}}}
\nc{\dott}{\;\cdot\;}

\def\0 {\nonumber}

\begin{document}
\setcounter{page}{0}
\newcommand{\inv}[1]{{#1}^{-1}} 
\renewcommand{\theequation}{\thesection.\arabic{equation}}
\newcommand{\be}{\begin{equation}}
\newcommand{\ee}{\end{equation}}
\newcommand{\bea}{\begin{eqnarray}}
\newcommand{\eea}{\end{eqnarray}}
\newcommand{\re}[1]{(\ref{#1})}
\newcommand{\qv}{\quad ,}
\newcommand{\qp}{\quad .}
\begin{titlepage}
\begin{center}

\hfill SISSA 43/2007/EP\\

\vskip .3in \noindent


{\Large \bf{Topological Gauge Theories on Local Spaces}\\
{\bf and}\\
\vspace{.3cm}
{\bf Black Hole Entropy Countings}} \\

\vskip .2in

{\bf Giulio Bonelli and Alessandro Tanzini}

\vskip .05in
{\em\small International School of Advanced Studies (SISSA) and INFN, Sezione di Trieste\\
 via Beirut 2-4, 34014 Trieste, Italy}
\vskip .5in
\end{center}
\begin{center} {\bf ABSTRACT }
\end{center}
\begin{quotation}\noindent
We study cohomological gauge theories on total spaces
of holomorphic line bundles over complex manifolds
and obtain
their reduction to the base manifold
by $U(1)$ equivariant localization of the
path integral.
We exemplify this general mechanism by proving
via exact path integral localization a reduction
for local curves conjectured
in hep-th/0411280, relevant to the calculation of black hole
entropy/Gromov-Witten invariants. Agreement
with the four-dimensional gauge theory is recovered by taking into
account in the latter non-trivial contributions coming from one-loop fluctuations
determinants at the boundary of the total space. We also study a
class of abelian gauge theories on Calabi-Yau local surfaces,
describing the quantum foam for the A-model,
relevant to the calculation of Donaldson-Thomas invariants.
\end{quotation}
\vfill
\eject

\end{titlepage}

\tableofcontents

\section{Introduction}

Topological theories are a natural and amusing instrument to quantify some non
perturbative aspects of superstring theory. Actually, BPS protection
reduces the evaluation of $F$-terms to a restricted configuration
space and this, in favorable conditions, allows their exact
calculation. The language of topological theories renders manifest
the non renormalization properties of such terms and clarifies the
above configuration space reduction \cite{agnt,bcov}. In particular, this applies
both to the world-sheet and the gauge theory approaches to
the counting of the entropy of BPS black holes in superstring
theories \cite{osv}. On the world-sheet side one finds that few, but
remarkable, all-loops calculations of the effective theory of 
a Calabi-Yau compactification can be
recasted in terms of amplitudes of a topological theory of strings
counting the number of inequivalent world-sheet instantons in the Calabi-Yau, 
$Z_{\rm top}=\sum_g \lambda_{\rm top}^{2g-2}F_{g}$. The
counting of D-brane bound states has been advocated in \cite{osv} to provide a
non-perturbative completion of these amplitudes via a conjectured
S-duality in topological strings \cite{nov}. A natural setting to describe this
duality is topological M-theory \cite{samson,m,bt,z,deBoer,B1,Anguelova}. 
More precisely, in \cite{osv} the suggestive relation $Z_{\rm BH}\sim |Z_{\rm top}|^2$, 
was proposed to hold in the limit of large black holes charges up to $O(e^{-N})$
corrections. 
In the dual D-brane language, the black hole BPS
multiplicities gets calculated by the supersymmetric partition
function of the twisted gauge theory living on the D-brane system.
It is therefore interesting to find exact calculational methods in
such a framework. In this paper we study in particular
cohomological gauge theories on local spaces, that is on spaces
whose geometry can be obtained by zooming in the vicinity of a given
non trivial cycle.
Namely, in the case of complex codimension one,
one has the total space over the cycle of an appropriate line
bundle. On this space, there is a natural $U(1)$ symmetry acting on
the fiber which can be used to simplify the relevant path integral
calculation. Actually, we will show how to realize the dimensional
reduction on the base by using such a symmetry. This will be
described in detail for the D4/D2/D0 system considered in \cite{vafa,aosv}.
We will also present a more general mechanism for the D6/D2/D0
system \cite{qf}.

In \cite{aosv} it has been proposed that the partition function
of the relevant twisted ${\cal N}=4$ gauge theory on the total space $\L\to\Sigma$,
where $\Sigma$ is a Riemann surface and $\L$ a line bundle over it,
can be evaluated by reducing to a q-deformed $YM_2$ on $\Sigma$ \cite{others}.
This was verified by comparing the factorized structure of the partition function in the large $N$ limit with
topological string amplitudes on the relevant Calabi-Yau local curve $\L\oplus \K\L^{-1} \to \Sigma$, where $\K$ is the
canonical line bundle on the base $\Sigma$ and the D4-branes wrap the cycle $\L\to\Sigma$ \cite{bryan}.

On the other hand, a comparison with the four dimensional gauge theory has been done in the genus zero case
when the instanton counting can be performed \cite{adlr,ffr}. In this case, agreement was found up to some perturbative contributions
which can be recasted as one-loop fluctuations of the Chern-Simons theory living at the boundary
of the total space \cite{adlr}. In this paper we suggest that these contributions have actually a very simple
interpretation in the four dimensional gauge theory. The crucial point is that we are quantising the theory
on a non-compact space. Henceforth, the one-loop determinants
do not cancel completely as it happens on compact manifolds, but they receive a
non-trivial contribution precisely from the boundary, where the four-dimensional action
reduces to a Chern-Simons term\footnote{In the case of flat $\bR^4$ space the boundary contribution is trivial.}.
By including these contribution in the four-dimensional instanton counting 
one can recover agreement with the q-deformed $YM_2$ results.

Let us briefly outline  the content of the paper.
In section 2 we start describing the geometric set-up, 
then we obtain the reduction of the topological action on the base
as well as the natural implementation of the $U(1)$-equivariant BRST symmetry in 
our problem.
In section 3 we analyze in close detail the D4/D2/D0 system and we discuss its reduction to the q-deformed $YM_2$ by using the relevant
localization of the path integral.
In section 4 we extend our procedure to the case of a topological gauge theory on a local Calabi-Yau surface and we propose an analogous
reduced approach to the counting of D6/D2/D0 BPS black hole entropy.
In section 5 we discuss few open issues.

\section{Reduction of cohomological gauge theories on local spaces}
\label{redux}

\subsection{Cohomological gauge theories on local spaces}

Let $\Sigma$ be a complex manifold with $dim_{\bf C}\Sigma=n$ and $\L$ be an holomorphic line bundle over it.
Let $M=\L\to \Sigma$ be the total space of $\L$.
Let $E$ be a gauge bundle over $\Sigma$ with structure group $G$. It extends canonically to a gauge
bundle on $M$ that we will still call $E$ for notational simplicity.
Let $A\in Conn(E,M)$ be a connection of $E$ on $M$ and let us consider a topological invariant
functional $S_{top}(A)$, that is invariant under continuous deformations of the connection
$A$ along $Conn(E,M)$. Typically
\be
S_{top}(A)=\int_M \left[P(F)\wedge K\right]
\label{ta}\ee
where
$P(F)$ is an $Ad-$invariant polynomial in the curvature $F=dA+A\wedge A$
and
$K\in H^\bullet(\Sigma)$ is an element in the even cohomology of $\Sigma$.
In (\ref{ta}) the top component of the integrand is understood.

This kind of theories naturally arise in D-brane/black holes entropy countings.
These countings reduce to the evaluation of the partition function of
the cohomological gauge theories obtained by quantizing (\ref{ta}).
Actually, since $M$ is non compact, we wish to sum over all
the gauge field configurations at the boundary.
This can be achieved either directly in the path integral by
implementing the summation over the boundary fluctuations via the associated Chern-Simons-like theory,
or by analyzing the sum over the boundary values in terms of the holonomy of gauge field at infinity.
The first approach leads then to work on the total space $M$ giving the result in terms of bulk and boundary
contributions. The second, as we will show in detail later, leads to a reduction of the calculation
of the path integral via an associated topological theory defined on the base $\Sigma$.

By construction $\partial M$ is the
total space of the circle bundle $arg(\L)$, namely $\partial M= arg{\L}\to\Sigma$.
Therefore, to parametrize the boundary conditions we can specify the holonomy
of $A$ along each $S^1$ fiber on $\Sigma$, that is
\be
e^{i\Phi}= P{\rm exp}\left\{i\int_{S^1\times Pt} A\right\}
\label{holonomy}
\ee
at any point $Pt\in\Sigma$. Summing over all the boundary conditions, then will mean path integrate over
the holonomy field $\Phi$.

On $M$ it is natural to distinguish horizontal and vertical directions.
Let $\gamma\in Conn(\L,\Sigma)$ be a reference connection for $\L$ and let $w$ be a coordinate
on the fiber. Then, the differential element $Dw=dw+\gamma w$ is covariant, namely
$Dw$ transform like a section of $\L$.
This defines a decomposition of the Dolbeault differential $\partial=\partial^h+\partial^v$, where
$$\partial^h=\partial_\Sigma - \gamma w\partial_w$$
$$\partial^v=Dw\partial_w$$
and $\partial_\Sigma$ is the Dolbeaux differential on the base.
The de Rham differential gets decomposed too.
Similarly, the gauge connection splits as
\be
A=A^h+A^v=
A^h+\varphi Dw+\varphi^\dagger D\bw
\label{split}\ee
where both $A^h\in Conn(E,\Sigma)$ and $\varphi\in\Gamma(Ad E,\Sigma)$
depend parametrically on the fiber coordinate $w$.

The above holonomy assignment (\ref{holonomy}) corresponds, up to a gauge transformation,
to the boundary conditions
\be
i_\theta A= i(w\varphi-\bw\varphi^\dagger)
\sim \frac{\Phi}{2\pi}\quad {\rm as}\quad w\sim\infty
\label{bc}\ee
where $\theta=i(w\partial_w-\bw\partial_\bw)$ is the fundamental vector field generating the
$U(1)$ action on the fiber $w\to e^{i\theta}w$.

We will now show how the topological theory (\ref{ta}) can be calculated in terms of reduced field configurations
on the base manifold. We will first analyse (\ref{ta}) at the classical level, and then show how
the equivariant localization with respect to the $U(1)$ action $w\to e^{i\theta} w$
on the fiber allows us to dimensionally reduce the theory at the path integral level.

\subsection{Covariant dimensional reduction of the classical action}
\label{covdim}

Since the classical topological action is independent upon the specific connection
of the given vector bundle $E$ and at given boundary conditions
we use to calculate it, we can chose to evaluate it by using some particular class
of connections, namely $U(1)$-invariant ones.
On top of it, to simplify our life,
we can perform the calculation in a given gauge with respect to the $G$ bundle. We
choose to work in radial gauge
$i_RA=0$, where
$R=w\partial_w+\bw\partial_\bw$ is the invariant vector field generating the fiber 
dilatations.
This means that we consider gauge connections of the form
\be
\hat A=\alpha+ \frac{\rho}{2\pi}{\rm Im}\left(\frac{Dw}{w}\right)
\label{iuuan}\ee
where $\alpha\in Conn(E,\Sigma)$. $\alpha$ and $\rho$ do not depend on ${\rm arg}(w)$ and
$\rho\in Adj(E,\Sigma)$ satisfies the boundary condition $\rho\sim\Phi$ as $w\sim\infty$.

Although the calculation can be performed in arbitrary dimensions, let us do it explicitly
in the case $n=1$ first. The relevant possible terms in (\ref{ta}) are $\int_M Tr(F\wedge F)$ and
$\int_M K\wedge Tr(F)$. By simply substituting (\ref{iuuan}), using Stokes theorem and the
boundary condition for $\rho$, we get
\be
\int_M tr \hat F\wedge K=\int_\Sigma tr\Phi K
\label{uno}
\ee
and
\be
\int_M tr(\hat F\wedge \hat F)=
\int_\Sigma \left[2tr\left(\Phi f \right) + \frac{1}{2\pi} tr\Phi^2 R_\L \right]
\label{due}\ee
where $f=d_\Sigma a + a\wedge a$, $~a=\alpha|_{w\sim\infty}$ and $R_\L= d_\Sigma {\rm Im}\gamma$
is the curvature of $\L$.

In the general case\footnote{Actually one could proceed also in the direct way.
This implies the use of the standard integral
transgression formula to solve $P(F)=d Q(A,F)$, passing to the boundary via Stoke's theorem,
calculating the integral along the circle at infinity and finally integrating back the
transgression formula.}
case we can proceed by using equivariant localization under the $U(1)$ invariance in
the evaluation of the topological invariant (\ref{ta}).
The equivariant extension of the gauge curvature at the fixed point (\ref{iuuan}) is
given by
\be
\tilde F = \hat F + \mu
\label{bismunt}
\ee
where $\mu$ is the moment map of the $U(1)$-action, $d_{\hat A} \mu = - i_\theta \hat F$.
By inspection, one gets $\mu=\frac{\rho}{2\pi}$.
The localisation formula \cite{bgv} then gives
\be
\int_M \left[P(F)\wedge K\right] = \int_M \left[P(\tilde F)\wedge K\right]
= 2\pi \int_{\Sigma} \left[\frac{P \left(f + \frac{\Phi}{2\pi}(1+ R_\L)\right)\wedge K}
{\left( 1 + R_\L \right)}\right]
\label{modo}
\ee
where the factor $\frac{2\pi}{1+R_\L}$ is the inverse Euler class of the normal bundle,
that is of the line bundle $\L$ itself.
Notice that since $K\in H^\bullet(\Sigma)$
then $i_\theta K =0$ and $K$ is trivially equivariantly closed.
Obviously, by specifying the general formula (\ref{modo}) to $n=1$, we get by expanding it
(\ref{uno}) and (\ref{due}).

Therefore, we see that if we constrain the connection to be $U(1)$ invariant, then the topological
action (\ref{ta}) gets reduced to (\ref{modo}) which depends on a connection for $E$ on the base
and the holonomy field.
Our next aim will be to implement the above reduction in the full gauge-fixed topological theory.

\subsection{Reduction via $U(1)$--equivariant localization: the four dimensional case}
\label{four}

To implement the reduction of the previous section at the path integral level
we use equivariant localisation with respect to the lifting on the field space of the
$U(1)$ action on the fiber $w\to e^{i\theta} w$, $\theta \in S^1$.

Before doing it, let us discuss some aspects of the topological gauge symmetry at hand.
We restrict for simplicity to the four dimensional case ($n=1$), although the following
strategy can be generalized to higher dimensions. This will be exemplified later for $n=2$.

Let us consider first the topological action $\frac{1}{2}\int_M tr F\wedge F$.
This is invariant under the BRST action on the space of connections with arbitrary boundary conditions
$sA=\Psi$ if $\Psi\sim d_A c$ at the boundary.
Actually, this is the usual scheme to quantize the four dimensional topological theory and to show
that it localises on instanton solutions. As it is well known the semiclassical limit is exact in these
cohomological theories and the path integral is reduced to the integration of the relevant observables
over the instanton moduli space times the contribution of the determinants of the one-loop fluctuations
around the instanton vacua. These determinants usually cancel to one due to supersymmetry.
Notice however that there is a subtlety in this case due to the non-compactness of the manifold
on which we are quantizing the theory. The functional integration of the fields at the boundary
produces a non trivial one-loop determinant which has to be taken into account.
Then schematically one gets
\be
Z_M= Z_{\partial M}^{CS(1-loop)} Z_M^{instantons}
\label{quattrod}\ee
the one-loop Chern Simons partition function comes from the integration along
the field fluctuations at the boundary while the second factor comes from the bulk
instantons.

The same calculation can be performed in a different way
by parametrizing the boundary values of the connection
via its holonomy. So we calculate
\be
Z_M=\int D\left[\Phi\right] Z_M(\Phi)
\label{dued}\ee
where $Z_M(\Phi)$ is the partition function calculated at fixed holonomy $\Phi$.
As we will show in detail in the next section, by applying this procedure
to the gauge theory relevant for the D4/D2/D0 system one gets the
partition function of the q-deformed $YM_2$.
The quantum measure $D\left[\Phi\right]$ has been discussed in \cite{aosv}
and we will come back to it later on. Notice that the explicit calculations
of the q-deformed $YM_2$ on the sphere precisely reproduce the structure in
(\ref{quattrod}) \cite{adlr}.

Let us put again the question about the boundary conditions on gauge parameters.
Now, since the holonomy is fixed, we can assign them differently.
Actually, at fixed holonomy, we require the natural boundary conditions
\bea
d_\theta A^h \sim 0  && i_\theta A \sim \Phi/2\pi \\
d_\theta \Psi^h \sim 0  && i_\theta\Psi\sim 0
\label{yess}\eea
up to gauge transformations, which we want to insist on.
Actually the topological action is not shift symmetric under the above boundary conditions.
In fact we have
\bea
s\frac{1}{2}\int_M tr F\wedge F &=& \int_M tr F\wedge d_A \Psi
= \int_{\partial M} tr \left( F^{hh} \wedge\Psi_\theta +  
F^{h}_{\theta}\wedge \Psi^h \right)
\nonumber \\
&=& \int_{\partial M} tr d_{A^h} A_\theta\wedge \Psi^h
= \int_{\Sigma} tr d_a \Phi \wedge\psi
\label{noninv}
\eea
where $A^h\sim a$ and $\Psi^h\sim\psi$ as $w\sim \infty$. Notice that, if $K$ is a 2-form on $\Sigma$, we have
$$
s\int_M K \wedge tr F=\int_{\partial M} K\wedge \Psi=0
$$
since $\Psi_\theta\sim 0$.

In order to cure the above lack of symmetry, one can improve the topological action
by adding the following topological observable localized on $\Sigma$
\be
{\cal O}^{(2)}_\Sigma = \int_\Sigma tr \left(\phi F + \frac{1}{2} \Psi\wedge\Psi\right)
\label{obs}
\ee
Let us read now the above deformation from the point of view of the calculation (\ref{dued})
with fixed holonomy at infinity.
We just noticed in (\ref{noninv}) that we need to improve our action.
In fact we have to improve the BRST symmetry too.
Let us consider the quantity $I=\frac{1}{2}\int_M Tr F\wedge F+ {\cal O}^{(2)}_{\Sigma_\infty}$
where $\Sigma_\infty$ is the copy of the base at infinity,
and calculate its BRST transform under a BRST operator $s_\theta$ to be determined
up to its fixed shift action on the connection $s_\theta A=\Psi$.
We have
\be
s_\theta I = \int_{\Sigma_\infty} tr (s_\theta\phi F) + tr\left(s_\theta\Psi - d_A\phi -d_A \Phi \right)\Psi
\ee
which is zero for
\be
s_\theta A=\Psi\quad , \quad
s_\theta\Psi= d_A(\phi+2\pi i_\theta A) + 2\pi i_\theta F
\quad {\rm and}\quad
s_\theta\phi=0
\label{preequiv}\ee
due to the boundary condition $i_\theta\Psi\sim 0$ at infinity.
Actually, we can rewrite (\ref{preequiv}) as
\be
s_\theta A = \Psi \, , \, \quad s_\theta \Psi = d_A\phi + 2\pi L_\theta A \, \, ,
\quad {\rm and} \, \quad s_\theta \phi = 0
\label{equi1}
\ee
which satisfies
$s_\theta^2 = \delta_\phi + 2\pi L_\theta$, where
$L_\theta = d i_\theta + i_\theta d$ is the Lie derivative along the vector
field $\theta$.
This reveals $s_\theta$ to be nothing but the $U(1)$-equivariant extension of the
original BRST symmetry.

Alternatively, by redefining
\be
\phi' = \phi + 2\pi i_\theta A
\label{schift}
\ee
we can rewrite (\ref{equi1}) as
\be
s_\theta A = \Psi \, , \, \quad s_\theta \Psi = d_A \phi' +2\pi i_\theta F \, \, ,
\quad {\rm and} \, \quad s_\theta \phi' = 2\pi i_\theta \Psi
\label{quelladopo}
\ee
and we have $s_\theta^2 = \delta_{\phi'} + 2\pi\L_\theta$, where
$\L_\theta = d_A i_\theta + i_\theta d_A$ is the {\it covariant} Lie derivative.
Notice that the field redefinition (\ref{schift}) changes the boundary
conditions of the field $\phi$ due to (\ref{bc}).
Actually these are no longer vanishing, but $\phi'\sim \Phi$ as $w\sim\infty$.

For later use, let us notice also that, due to (\ref{yess}), we have $s_\theta \int_M K\wedge tr F=0$.

\section{D4/D2/D0 branes on local Calabi-Yau's}

Let us here consider the topological gauge theory studied in
\cite{vafa}, \cite{aosv}.
Consider a Riemann
surface $\Sigma$, a line bundle $\L$ over it and the CY local curve
$X=\L\oplus \L^{-1} \K \to \Sigma$, where $\K$ is the canonical line
bundle over $\Sigma$. Let us place $N$ D4-branes on $M=\L\to\Sigma$
and any number of D0 and D2 wrapping $\Sigma$.

Actually, the entropy of such a D-brane system is calculated by the
Vafa-Witten twisted ${\cal N}=4$ SYM with gauge group $U(N)$ and
classical action \be S_{top}=\frac{1}{2g_s}\int_M tr(F\wedge F) +
\frac{\theta}{g_s}\int_M tr F \wedge K \label{aosv} \ee with $K$ the
unit volume form on $\Sigma$.

In the previous section we have been proving that the path integral for such a theory
gets localized on $U(1)$-invariant configurations.
In particular, the reduction of the various terms in (\ref{aosv})
reads
\be
S_{red}=\frac{1}{g_s}\int_\Sigma tr\left(\Phi f \right) +
\frac{\theta}{g_s}\int_\Sigma tr\Phi K +\frac{1}{2 g_s}\int_\Sigma
tr\Phi^2 \frac{R_\L}{2\pi}
\label{saosvred}
\ee
Let us notice the
appearance of the last term quadratic in $\Phi$ which we obtained
just by insisting on a covariant reduction scheme.
The curvature of the line bundle is $R_\L= - 2\pi p K$, where \break $deg(\L)=-p$.
In this way one obtains precisely the topological action for the q-deformed $YM_2$ considered in
\cite{aosv}.

Let us now discuss the evaluation of the partition function for the
D4/D2/D0 system
\be
Z_{D4/D2/D0} = \int {\cal D}[A,\Psi \ldots] {\rm exp}\left[-S_{top} - S_{g.f.} - \frac{1}{g_s}\left({\cal O}_2 +
\theta {\cal O}_2^{[K]} + {\cal O}_4 \right)\right]
\label{part4}
\ee
where
\bea
{\cal O}_2 &=& \int_\Sigma tr \left(F \phi + \frac{1}{2}\Psi\wedge\Psi\right) \\
{\cal O}_2^{[K]} &=& \int_\Sigma tr\phi K \\
{\cal O}_4 &=& \frac{1}{2}\int_\Sigma tr\phi^2 R_\L
\label{obs1}
\eea
The subscript index in (\ref{obs1}) is the ghost number.

Notice that, according to the prescription described in Sect. 2.3,
in (\ref{part4}) we inserted the topological observables
(\ref{obs1}) with positive ghost numbers. Due to the absence of a
ghost number anomaly in the twisted ${\cal N}=4$ SYM theory, the
v.e.v. of these observables is actually vanishing and the path
integral is insensitive to their presence. Nonetheless, their
insertion simplifies the task of dimensional reduction. This
procedure is completely analogous to the mass deformation of the
Vafa-Witten theory which allows to compute the twisted ${\cal N}=4 $
partition function in terms of ${\cal N}=1$ vacua countings. In
our case the observables (\ref{obs1}) break the anti-BRST of the
$N_T=2$ balanced topological field theory \cite{BT,DM} thus reducing the twisted
supersymmetry to $N_T=1$. Indeed, the deformation (\ref{part4})
suggests an inspiring relation with Donaldson intersection theory on
the instanton moduli space. It would be very interesting to study
in deeper detail this connection by analysing the $U(1)$-localisation
procedure for Witten's TYM theory \cite{tym}. For Riemann surfaces
of genus zero, this could provide a direct derivation of 
the blow-up formulae for Donaldson polynomials \cite{fs} that were
obtained in \cite{WM,MMs} by using the low-energy Seiberg-Witten
effective theory.

At the fixed locus under the $U(1)$-action, the observables (\ref{obs1}) give
\bea
{\cal O}_2^{red} &=& \int_\Sigma tr \left((f+R_\L\Phi)\phi + \frac{1}{2}\psi\wedge\psi\right) \\
{\cal O}_2^{[K]\,red} &=& \int_\Sigma tr\phi K \\
{\cal O}_4^{red} &=& \frac{1}{2}\int_\Sigma tr\phi^2 R_\L
\label{redobs}\eea
where all the fields depend on the base $\Sigma$ only.
We calculate the total classical action first
by reducing $S_{top}+\frac{1}{g_s}\left[{\cal O}_2+\theta{\cal O}_2^{[K]}
+{\cal O}_{4}\right]$ and obtain
\be
S_{red}=
\frac{1}{g_s}\int_\Sigma \left[\theta tr(\phi')K + tr(\phi'f+\psi^2)
+\frac{1}{2}tr\left({\phi'}^2\right)R_\L\right]
\label{2dredaction}\ee
where $\phi'=\phi+\Phi$. Notice that this is {\it precisely} the shift (\ref{schift}) so that
we can interpret the reduced theory as a two dimensional topological gauge theory on $\Sigma$.

Notice however that since we choose $\Sigma=\Sigma_\infty$, the
vanishing boundary conditions on $\phi$ imply that on the base image
at infinity $\phi'=\Phi$. Therefore, we get the reduced theory with
action
\be
S_{YM_{2}}= \frac{1}{g_s}\int_\Sigma \left[\theta tr(\Phi)K
+ tr(\Phi f+\psi^2) +\frac{1}{2}tr\left({\Phi}^2\right)R_\L\right]
\label{2dYMaction}
\ee
that is the topological action of the 2d Yang-Mills. The reduced BRST operator is then
\be
s a=\psi \quad ,
\quad s\psi = d_a \Phi \quad {\rm and} \quad s\Phi=0
\label{2dbrst}
\ee

\subsection{Equivariant localisation of the path integral}

In order to show the dimensional reduction at the path integral level,
we have to face the full gauge fixed theory.
The full field content of the Vafa-Witten twisted ${\cal N}=4$
theory is given by the $N_T=2$ connection multiplet $(A,\Psi_\pm,H)$, a selfdual
tensor multiplet $(B^+,\chi^+_\pm,H^+)$ and the Cartan multiplet \cite{DM}. The relevant
$U(1)$-equivariant BRST transformations on the selfdual multiplet
are
\be
s_\theta B^+ = \chi^+ \quad,\quad s_\theta \chi^+= [\phi',
B^+] + \L_\theta B^+
\label{eqb}\ee
The gauge fixed action of the theory is given in terms of the action potential \cite{DM}
\bea
\F &=& \F_1+\F_2 \nonumber\\
\F_1&=&\int_M Tr\left[ B^+_{\mu\nu}\left(F^{\mu\nu}+\left[ B^\mu_{\;\;\rho},B^{\rho\nu}\right]\right)\right]\nonumber\\
\F_2 &=& \int_M Tr\left[
\chi^+_+\wedge  \chi^+_- +\Psi_+\wedge *\Psi_-
\right]
\label{acpot}
\eea
where we do not consider the gauge fixing of the Cartan multiplet since it can be universally re-absorbed \cite{DM}.

Let us now show that the modes with non zero transverse momentum can be integrated out and that they do not
contribute to the topological partition function.
In order to show this, we split all field configurations in terms of zero and non-zero modes
with respect to the operator which defines the transverse momentum. Because
of the gauge symmetry of the problem we have to consider the
operator $d^v_{A_\infty}$, where $A_\infty$ is the connection at
the boundary $w\sim\infty$. The integrability of $F^{vv}$ along the
fibers requires that
$F^{vv}_{\infty}=\left(d^v_{A_\infty}\right)^2=0$.
To choose the appropriate gauge fixing,
let's rescale the fields as
\bea
A&=&A_\infty + x\delta A^h +x^{-1} \delta A^v,\nonumber\\
B^+&=& x^{-1}\delta B^{(2,0)} + c.c. +
(b_\infty + \delta b)\omega_x.
\label{rescalings}
\eea
where $\delta[fields]$ denote the non zero modes with respect to
$d^v_{A_\infty}$. We will consider the scaling gauge $x\to\infty$.
In (\ref{rescalings}) we used the rescaled K\"ahler form
$\omega_x=x\omega^{hh}+x^{-\alpha}\omega^{vv}$
obtained from
the block diagonal $\omega=\omega^{hh}+\omega^{vv}=i\partial\bar\partial f$, where $f$ is the K\"ahler potential.
Notice that in (\ref{rescalings}) we retained only the zero modes for the connection multiplet
and the K\"ahler component of the selfdual multiplet,
while we did not for the $(2,0)$ part of the selfdual multiplet.
Actually $B^{(2,0)}$ would have a zero mode structure
\be
B^{(2,0)}_\infty= \beta\frac{Dw}{w}
\label{binfinity}\ee
up to gauge conjugation by $e^{i\theta\Phi}$ and where $\beta\in T^{(1,0)}(\Sigma, Adj E)$.
This field configuration is singular as $w\sim 0$ and we do not consider it to belong to the allowed field space
\footnote{Notice that, if one would keep it, this sector would extend the $YM_2$ theory on $\Sigma$ to a deformed
Hitchin system with equations
$$
f+\Phi R_\L=[\beta,\bar\beta]
\quad ,\quad
\bar\partial_a \beta=0
\quad{\rm and}\quad
d_a\Phi=0$$
$$
[\Phi,\beta]=0
\quad ,\quad
d_a b=0.
$$}.

In (\ref{rescalings}) we just gave the
rescaling of each top member of the $N_T=2$ multiplets, the others
being equal because of BRST invariance. This guarantees that such a
rescaling doesn't generate any non trivial Jacobian from the quantum
measure in the path-integral.
The configuration $A_\infty$ is taken to
satisfy the holonomy assignment and is given, up to gauge
conjugation, by
$$A_\infty=a+\frac{\Phi}{2\pi}{\rm Im}\left(\frac{Dw}{w}\right).$$

To calculate the rescaled gauge fixing action,
we apply the rescalings (\ref{rescalings}) to (\ref{acpot}) and obtain that if $\alpha >2$ then
\bea
\F_1&=&\int_M \delta B^{(2,0)}\bar\partial^v_{A_\infty}(\delta A^{h})^{(0,1)} + c.c. + \delta b\omega^{hh}
d^v_{A_\infty}\delta A^v
+O(1/x)
\label{f1}\eea
while
$\F_2 = O(1/x)$
and does not contribute in the scaling limit.

In order to integrate the left over transverse momentum zero modes for the
BRST doublets $(\bar\psi, H)_\infty$ and $(b,\chi_+)_\infty$ $(\chi_-,h)_\infty$
we add to the gauge fixing action the exact term
$$
s_\theta\int_\Sigma tr\left(b[\Phi,\chi_-]+H\wedge *[\Phi,\bar\psi] \right)
$$
which path integrates to $det(Adj_\Phi)^2$ in the fermionic sector and to $det(Adj_\Phi)^{-2}$
in the bosonic one so giving no contribution to the partition function.

Therefore, in the scaling gauge $x\to\infty$ the left over term (\ref{f1}) gauge fixes all the fluctuation in the connection
and in the selfdual $N_T=2$ multiplets.

Summarizing, in the scaled gauge we then find that all the modes with non vanishing transverse momentum get gauge fixed to zero
and the path integral evaluation reduces to the theory dimensionally reduced on the basis with action
(\ref{2dYMaction}). Notice that, due to the compactness of the holonomy field $\Phi$ its path integral measure
has to be suitably defined as in \cite{aosv} therefore giving the q-deformation of $YM_2$ on $\Sigma$.

\section{Quantum foam on the local CY surface}
\label{quantumf}

Let us consider in this section a topological abelian $U(1)$ gauge theory on a generic
CY local surface $M=\L\to \Sigma$, where
$\L=\K$ is the canonical line bundle of the complex surface $\Sigma$ itself.
This theory describes a system of D2/D0 dissolved in a D6 brane wrapping $M$.
The topological partition function of this theory computes the Donaldson-Thomas 
invariants of the CY it is defined on.
Moreover, it has been proposed in \cite{qf} as weak coupling expansion
in $g_s$ for the topological A-model.
Our reduction formula works also if the line bundle $\L$ is not constrained to be $\K$
by the Calabi-Yau condition.

The quantum foam topological action is
\be
S_{qf}(A)= \frac{g_s}{3}\int_M F\wedge F\wedge F +\int_M k_0\wedge F\wedge F
\label{foam}\ee
where $F=dA$ is the abelian curvature and $k_0\in H^2(\Sigma,Z)$. By definition, the topological
version to consider is the twisted maximally supersymmetric one.

As in the four dimensional case, we can consider the quantization of the theory from two
equivalent points of view.
Its partition function can be evaluated directly, by giving
\be
Z_{qf}=Z^{(5d - one loop)} Z_{bulk}
\label{qfsix}\ee
where $Z^{(5d - one loop)}$ is the one loop path integral of the topological boundary theory
with action
\be
S_{5d}(A)=\frac{g_s}{3}\int_{\partial M} A\wedge dA\wedge dA +\int_M k_0\wedge A\wedge dA
\label{bt}\ee
The bulk contribution $Z_{bulk}$ is the partition function of the topological theory
on $M$ with vanishing boundary conditions.

A second equivalent calculational scheme is again obtained by specifying the (abelian) holonomy
$e^{i\phi}=e^{i\int_{S^1_\infty} A}$.
Therefore we have also
\be
Z_{qf}=\int {\cal D}[\Phi] Z_{qf}(\Phi)
\label{qffour}\ee

By applying the same procedure we developed in the previous sections, we can now
reduce the evaluation of the partition function to that of an abelian
topological gauge theory on the four dimensional base manifold.

The only peculiar ingredient to be fixed is the observable of the initial theory
to be added in order to produce the equivariant extension of the action functional.
This is given by a suitable linear combination of
\be
{\cal O}_1=\int_\Sigma \Psi\wedge\Psi\wedge F +\phi F \wedge F
\quad{\rm and}\quad
{\cal O}_2=\int_\Sigma k_0\wedge \left(\frac{1}{2}\Psi\wedge \Psi + \phi F\right)
\label{sixobs}\ee

We can calculate then, by using (\ref{modo})
$$
\int_M \tilde F\wedge \tilde F\wedge \tilde F=
\int_\Sigma\left[
\Phi^3 \frac{R_\K}{2\pi}\wedge\frac{R_\K}{2\pi}
+ 3 \Phi^2 f \wedge \frac{R_\K}{2\pi}+ 3 \Phi f \wedge f
\right]
$$
and
$$
\int_M k_0\wedge \tilde F\wedge \tilde F=
\int_\Sigma
k_0\wedge\left[2 f \Phi +\frac{R_\K}{2\pi} \Phi^2\right]
$$
and we get that the quantum foam reduces to the topological action on the base $\Sigma$
\bea
S_{qf-4d}&=&g_s\int_\Sigma \left[
\frac{1}{3}\Phi^3\frac{R_\K}{2\pi}\wedge\frac{R_\K}{2\pi}+
\Phi^2 f \wedge \frac{R_\K}{2\pi}+
\Phi f\wedge f
+\left(f+\Phi\frac{R_\K}{2\pi}\right)\wedge\psi\wedge\psi
\right]
\nonumber\\
&&+\int_\Sigma k_0\wedge\left(\Phi f+\Phi^2 \frac{R_\K}{2\pi}+\frac{1}{2}\psi\wedge\psi\right)
\label{qfred}\eea
that is closed under the reduced BRST action $sa = \psi$, $s\psi=d\Phi$ and $s\Phi=0$.
The maximally susy twisted theory that we are considering is the one studied
in \cite{park} on K\"ahler threefolds.
One can extend in full analogy to this case too the off shell localization of the path integral
that we discussed for the Vafa-Witten theory in the previous section.
Due to the abelian nature of the theory, the quantum measure over $\Phi$ is the usual translational invariant
one for a single $S^1$ valued field. 
We would get an analogue of the q-deformation in a non abelian version of the
topological theory at hand.
Therefore, we find that the calculation scheme at fixed holonomy
reduces to the cohomological gauge theory (\ref{qfred})
$$
Z_{qf}=\int {\cal D}\left[a,\psi,\Phi\right] e^{-S_{qf-4d}} \ \ .
$$
It is tempting to treat the theory defined by (\ref{qfred}) in a perturbative expansion
in $g_s$. Indeed, the propagators for this theory would be given by the terms in the
second line of (\ref{qfred}) and as such would be localised on two-dimensional cycles.
Moreover, the interaction terms in the first line give rise to cubic vertices.
This structure is similar to the one found for the topological vertex computing
Gromov-Witten invariants on toric CY's \cite{tv} and their generalisation
to degenerated torus actions studied in \cite{duiliu}. It would be nice to try
to make a closer comparison with these formalisms.

\section{Conclusions and open issues}

In this paper we have been developing a general reduction scheme for cohomological
gauge theories on local spaces. 
Actually we have been studying in detail the case of twisted maximally supersymmetric
gauge theories, but we believe that a more general reduction scheme holds also
for other cases. In particular a tempting relation among 
Donaldson invariants and two-dimensional Yang-Mills theories would appear
in the reduction of the twisted ${\cal N}=2$ theory.
Let us here discuss few open issues raised by the analysis we performed in this paper.

As far as the comparison among the calculation of the D4/D2/D0 partition function in the 
scheme in four and in two dimensions is concerned, it remains to fully recognize 
the precise encoding by the $qYM_2$ of the bulk point-like instantons \cite{ffr,adlr}.
Actually our reduction mechanism gives non-trivial results only for gauge fields
with non-trivial holonomy at infinity, suggesting the possible existence of a further
branch.
This could be related to the fact that 
in the vicinity of the base manifold we assumed some regularity 
conditions for the fields, see Sect.3.1. 
It is possible that our criteria are too strict and that we 
should include more general field configurations.
In particular the inclusion of singular fields via a suitable
compactification of the space of $U(1)$-invariant configurations could provide
the missing terms. From the analysis of \cite{Naka} one could expect that at least in the 
rank one case this compactification
contains the symmetric product $\Sigma^{[k]}$ of $k$ copies of the base manifold $\Sigma$,
$k$ being the instanton number. This would properly take into account the contribution
of point-like instantons.

In section \ref{quantumf} we generalized the dimensional reduction of the path integral
to the six- dimensional abelian gauge theory on the local CY surface $\K\to \Sigma$.
One should be able to compare the calculation of the partition function of the
reduced gauge theory on the four manifold $\Sigma$ with the topological 
string partition function on the local Calabi-Yau, for example in the 
case $\Sigma=\C P^2$ and $\K={\cal O}(-3)$.

The ability to show that the fields with non zero vertical momentum do not really contribute 
to the topological theory could be extended in other cases.
The mechanism we have shown could generalize to Schwarz--type theories, like Chern--Simons and 
holomorphic Chern--Simons giving an off-shell reduction for the constructions made in \cite{ricco}.
Moreover, it could extend to cohomological gauge theories on more general local spaces 
with higher number of non compact directions.

\noindent{\bf Acknowledgement}:
Let us thank D.E. Diaconescu, H. Ooguri and G. Thompson for useful discussions.
The authors wish to thank the Galileo Galilei Institute for Theoretical Physics for hospitality
during the last stages of this work.
The research of G.B. is supported by
the European Commission RTN Program MRTN-CT-2004-005104 and by MIUR.

\end{document}